\input{aipcheck}

\documentclass[
 ,final            
  ]
  {aipproc}

\layoutstyle{8x11single}
\begin{document}

\title{On possible origins of power-law distributions}


\author{Grzegorz Wilk}{
  address={National Centre for Nuclear Research, Ho\.za 69,
00-681 Warsaw, Poland\footnote{e-mail: wilk@fuw.edu.pl}} }

\author{Zbigniew W\l odarczyk}{
  address={Institute of Physics, Jan Kochanowski University,
\'Swi\c{e}tokrzyska 15, 25-406 Kielce, Poland\footnote{e-mail:
zbigniew.wlodarczyk@ujk.edu.pl}} }

\begin{abstract}
Selected examples of possible origins of power-law distributions
are presented.
\end{abstract}

\maketitle

\section{Introduction}

Power-law distributions are ubiquitous in all branches of science
and their possible origins are accordingly numerous. Here we
present and shortly discuss some selected examples chosen from our
experience with high energy multiparticle production processes
\cite{WWexample1,WWexample2,WWexample3}. Such processes are
usually described using some kind of thermodynamical approach with
Boltzmann-Gibbs (BG) statistics (or based on Shannon entropy if
treated from the information theory point of view). That was
because all important measured distributions apparently seemed to
follow exponential laws,  $f(x) \propto \exp(-x/x_0)$. However, it
was realized that, in fact, almost all of them are in fact
developing  power-like tails for large $x$, $f(x) \propto
x^{-\gamma}$, remaining exponential at small $x$. It was then
natural to propose a Tsallis distribution,
\begin{equation} f_q(x)\propto \left[1 - (1-q)\frac{x}{x_0}\right]^{\frac{1}{1-q}} =
\exp_q\left(-\frac{x}{x_0}\right), \label{eq:T}
\end{equation}
which possess both these features, as a most natural candidate for
the description of such processes (notice that for $q \rightarrow
1$ $\exp_q\left(-x/x_0\right)$ becomes the usual
$\exp\left(-x/x_0\right)$). This is also supported by the argument
that formula (\ref{eq:T}) follows from the replacement of BG
statistics by Tsallis statistics (based on Tsallis entropy).

On the other hand, formula (\ref{eq:T}) also describes data which
are not likely to follow a thermal approach but rather come from
some kind of hard collisions (described by, for example, quantum
chromodynamics, QCD) \cite{CYWWexample}. In such cases
justification of Eq. (\ref{eq:T}) must be different (cf.,
\cite{CYWWexample}). Therefore the search for other than
thermodynamical origins of Tsallis distribution power-laws is very
desirable. Among many possibilities (cf. \cite{WWexample1} for
more details and references) the most interesting from our point
of view is the so called {\it superstatistics}, {\it stochastic
network approach}, which we shall mention shortly, and connection
with {\it multiplicative noise} which will be discussed in more
detail below. We end with a short demonstration that essentially
all distributions of interest, including Tsallis, can be derived
from information theory based on Shannon entropy.

\section{Examples of mechanisms leading to Tsallis distribution}

{\bf Superstatistics} \cite{Ss}. It is based on the observation
that fluctuations of the scale parameter $\tilde{x}_0$ in
exponential distribution, for example given by a gamma function
\cite{WW,BP},
\begin{equation}
g(\tilde{x}_0) =
\frac{1}{\Gamma\left(\frac{1}{q-1}\right)}\frac{x_0}{q-1}
\left(\frac{1}{q-1}\frac{x_0}{\tilde{x}_0}\right)^{\frac{2-q}{q-1}}
\exp\left(-\frac{1}{q-1}\frac{x_0}{\tilde{x}_0}\right),
\label{eq:Gamma}
\end{equation}
convert $\exp\left(-x/x_0\right)$ into $\exp_q\left(-x/x_0\right)$
with parameter $q = 1 + Var(\tilde{x}_0)/\langle
\tilde{x}_0\rangle^2
> 1$ characterizing the strength of fluctuations.

{\bf Stochastic network approach} \cite{qnets}. This can be used
in situations when the scale parameter depends on the variable
under consideration. It is the case when in the system under
consideration one encounters correlations of the type called {\it
preferential attachment and "rich-get-richer" phenomenon} in
networks, for example,when $x_0 \rightarrow x'_0(x) = x_0
+(q-1)x$. In this case the probability distribution function,
$f(x)$, is given by equation the solution of which is Tsallis
distribution:
\begin{equation}
\frac{d f(x)}{dx} = \frac{1}{x'_0(x)} f(x) \qquad
\Longrightarrow\qquad f(x) = \frac{2 - q}{x_0} \left[ 1 - (1 -
q)\frac{x}{x_0} \right]^{\frac{1}{1 - q}}. \label{eq:netT}
\end{equation}

{\bf Tsallis distribution from multiplicative noise}
\cite{qnets,WWexample2}. Following ideas of \cite{BP}, consider
the Langevin equation
\begin{equation}
\frac{dp}{dt} + \gamma(t)p=\xi(t)  \label{eq:L}
\end{equation}
where both $\gamma(t)$ and $\xi(t)$ denote stochastic processes,
traditional multiplicative noise and additive noise, respectively.
The corresponding Fokker-Planck equation in this case is
\begin{equation}
\frac{\partial f}{\partial t} = - \frac{\partial \left(K_1
f\right)}{\partial p} + \frac{\partial^2\left( K_2
f\right)}{\partial p^2} \label{eq:FP}
\end{equation}
with coefficients
\begin{equation}
K_1 = E(\xi) - E(\gamma)p\qquad{\rm and}\qquad K_2 = Var(\xi) - 2
Cov(\xi,\gamma)p + Var(\gamma) p^2, \label{eq:K1K2}
\end{equation}
which stationary solution satisfies the equality
\begin{equation}
\frac{d\left(K_2 f\right)}{dp} = K_1 f.   \label{eq:eqKK}
\end{equation}
As shown in \cite{BP,AT} in the case of $Cov(\xi,\gamma) = 0$ and
$E(\xi) = 0$ (i.e., for, respectively, no correlation between
noises and no drift term due to the additive noise) the solution
of Eq. (\ref{eq:eqKK}) is given by the non-normalized Tsallis
distribution for the variable $p^2$:
\begin{equation}
f(p) = \left[ 1 + (q - 1)
\frac{p^2}{T}\right]^{\frac{q}{1-q}}\quad\quad {\rm with}\quad T =
\frac{2Var(\xi)}{E(\gamma)}\quad {\rm and}\quad q - 1 =
\frac{2Var(\gamma)}{E(\gamma)}\quad{\rm or}\quad T =
(q-1)\frac{Var(\xi)}{Var(\gamma)}. \label{eq:Tsallis2}
\end{equation}
However, we are interested in the Tsallis distribution
(\ref{eq:T}) with only a single power $p$, in particular in
whether such a distribution also satisfies the stationary equality
(\ref{eq:eqKK}). Write for simplicity (\ref{eq:T}) in the
following form:
\begin{equation}
f(p) = \left[ 1 + \frac{p}{nT}\right]^{-n},\qquad\qquad {\rm
where}\qquad n = \frac{1}{q - 1}. \label{eq:tn}
\end{equation}
We shall demonstrate now that Eq. (\ref{eq:tn}) satisfies
(\ref{eq:eqKK}) but under conditions which must be satisfied by
$T$ and $q$. Substituting (\ref{eq:tn}) into the stationary
equality (\ref{eq:eqKK}), one gets that $K_1$ and $K_2$ have to be
related in the following way:
\begin{equation}
K_2(p) = - \frac{nT + p}{n} \left[ K_1(p) -
\frac{dK_1(p)}{dp}\right]. \label{eq:Cond}
\end{equation}
Using definitions of $K_{1,2}$ given in Eq. (\ref{eq:K1K2}) one
finds that Eq. (\ref{eq:tn}) holds, provided
\begin{equation}
n = 2 + \frac{E(\gamma)}{Var(\gamma)}\quad{\rm and}\quad T =
\frac{1}{n(n-1)}\frac{E(\xi)}{2Var(\gamma)} - \frac{1}{n}
\frac{Cov(\xi,\gamma)}{Var(\gamma)} =
\frac{n-2}{n(n-1)}\frac{E(\xi)}{2E(\gamma)} -
\frac{n-2}{n}\frac{Cov(\xi,\gamma)}{E(\gamma)}. \label{eq:nT}
\end{equation}
In terms of the nonextensivity parameter $q$, one has
\begin{equation}
T=T(q)= - \frac{Cov(\xi,\gamma)}{E(\gamma)} + \left[
\frac{E(\xi)}{2E(\gamma)} + \frac{Cov(\xi,\gamma)}{E(\gamma)}
\right](q - 1) - \frac{E(\xi)}{2E(\gamma)} (q - 1)^2
=(2-q)\left[T_0+(q-1)T_1\right]
\label{eq:Tq}
\end{equation}
where
\begin{equation}
T_0 = - \frac{Cov(\xi,\gamma)}{E(\gamma)}\qquad {\rm  and}\qquad
T_1 = \frac{E(\xi)}{2E(\gamma)}. \label{eq:TEFF}
\end{equation}
It should be stressed that Eq. (\ref{eq:Tq}) provides a new
justification to the idea of {\it effective temperature},
\begin{equation}
T = T_0 - (q-1)T_0\quad (linear)\qquad {\rm or}\qquad  T = T_0 -
(q-1)^2T_1\quad (quadratic), \label{eq:i_2}
\end{equation}
\begin{figure}[h]
  \includegraphics[height=.29\textheight]{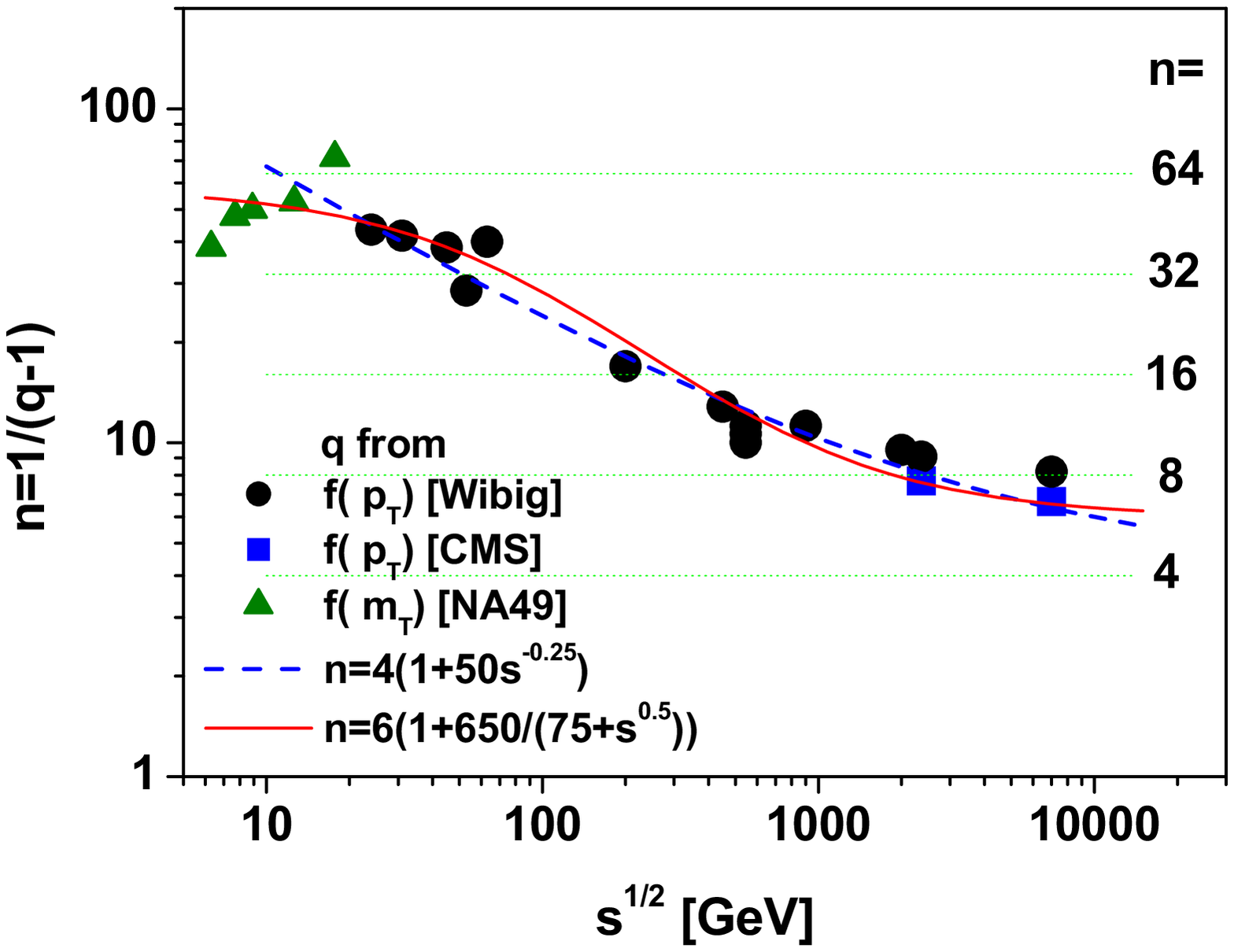}
  \includegraphics[height=.3\textheight]{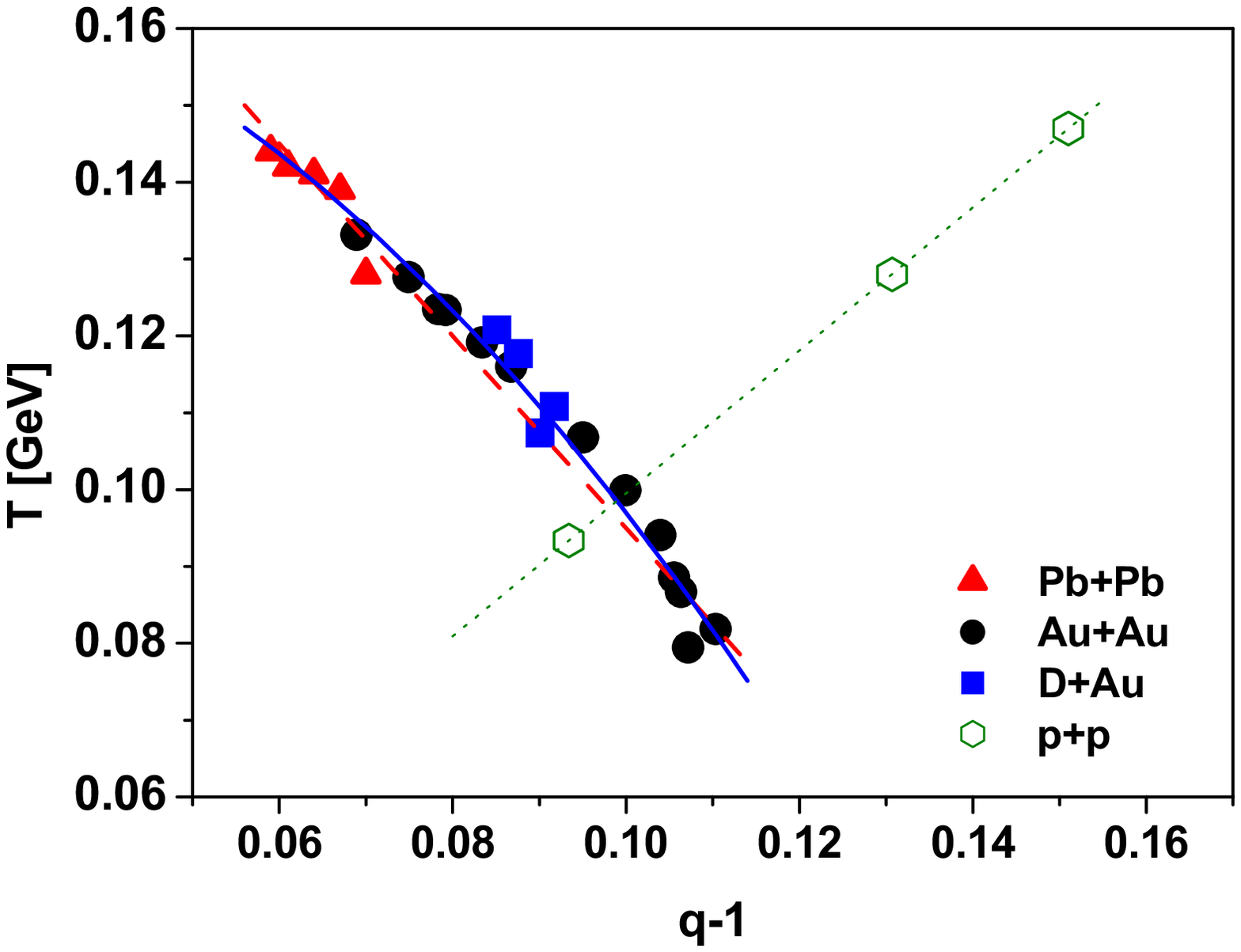}
  \caption{(Color online) Left panel: Energy ($\sqrt{s}$) dependence of the
  exponent $n=1/(q-1)$ (cf., Eq. (\ref{eq:nT}))deduced from NA49 \cite{NA49}
  and CMS \cite{CMS} experiments and from compilation \cite{Wibig} (Wibig).
  Two possible fits are shown. Right panel: $T$ as function of $(q-1)$
  (cf., Eq. (\ref{eq:Tq})) deduced from Pb+Pb collisions at energies
  $6.3,~7.6,~8.8,~12.3,~17.3$ GeV cite{NA49}, from Au+Au and D+Au
  collisions at energy $200$ GeV \cite{RHIC} and from p+p collisions at
  energies $200$ GeV \cite{RHIC} and $900$ and $7000$ GeV \cite{CMS,ATLAS}.
  The nuclear results can be fitted either by linear form,
  $T(q) = 0.22-1.25(q-1)$ (blue), or by quadratic, $T(q) = 0.17-7.3(q-1)^2$ (red).
  The best fit to p+p results is linear: $T(q) = 0.0065 + 0.93(q-1)$. }
  \label{Fig1}
\end{figure}
introduced by us in \cite{WWexample1}. Notice that if $E(\xi) =0$
(no drift due to the additive noise)the positivity of temperature
$T$ in Eq. (\ref{eq:nT}) requires that $Cov(\xi,\gamma) < 0$,
i.e., that one has an anticorrelation between noises. For $E(\xi)
\neq 0$ one can have $T > 0$ even for $Cov(\xi,\gamma)=0$.
However, in order to satisfy the experimentally observed linear
dependence of $T$ on $(q-1)$
\cite{WWexample1,WWexample2,WWexample3} one needs $Cov(\xi,\gamma)
< 0$ and $E(\xi) = 0$. For a quadratic dependence one can have
$E(\xi) \neq 0$. The examples of energy dependence of $n=1/(q-1)$
deduced from the available experimental data and $T$ as a function
of $q-1$ for different energies are presented in Fig. \ref{Fig1}.
Notice the opposite behavior of $T(q)$ for A+A and p+p collisions.
In the notation of Eq. (\ref{eq:Tq}) experimental data show that
in the presented range of $q$ for p+p collisions $E(\xi)
> 2|Cov(\xi,\gamma)|$ and also that for A+A collisions $E(\xi) <<
2|Cov(\xi,\gamma)|$ (and therefore it can become negative).
Actually, already in \cite{PRC} where we have analyzed $T(q) =
T_{eff}$ for the first time, the point for p+p for $200$ GeV was
not following the A+A results. In Fig. \ref{Fig1} the new points
from CMS p+p data at higher energies were added
\cite{CYWWexample}. It turned out that $T(q)$ for p+p collisions
are just opposite to that for A+A data \footnote{To explain this
difference one has to remind that behind $T_{eff} = T(q)$ used in
\cite{PRC} was the idea of energy transfer between the interaction
region and its surroundings. In the case of A+A collisions energy
can transferred to the spectator nucleons (not participating in
the collision process). Spectators, which have small transverse
energy, effectively "cool" the interacting system. For the p+p
collisions there is no such possibility. The region of interaction
is immersed in quark-gluon environment, which has transverse
energy comparable with that of the colliding system and because of
this they additionally "heat" it. As a result, the corresponding
parameter describing this energy transfer is negative for A+A
collisions and positive for p+p collisions, resulting in behavior
seen in Fig. \ref{Fig1}. However, before reaching any final
conclusions, one has to stress the difficulties in evaluation of
parameters $T$ and $q$ from experimental data. Among other things
both quantities depend on which phase factors were taken into
account and whether $m_T-m$ scaling was assumed or not ($T$ is
more sensitive to data at low transverse momenta, while $q$ to
higher transverse momenta points). Therefore more detailed
analysis are needed and are in progress.}.

{\bf Tsallis distribution from Shannon entropy}. In information
theory applications one uses Shannon entropy,
\begin{equation}
S = - \int dx\, f(x)\ln[f(x)], \label{eq:S}
\end{equation}
as a measure of information from which one can deduce different
probability distributions imposing different conditions. For
example, for condition $\langle x\rangle = const$ one gets the
usual exponential distribution. Adding to it condition $\langle
x^2\rangle = const$ results in a Gauss distribution whereas for
$\langle \ln(x)\rangle = const$ and $\langle \ln \left(1 +
x^2\right)\rangle = const$ one gets, respectively, gamma and
Cauchy distributions. In general, for some function of $x$,
$h(x)$, the maximum entropy density for $f(x)$  satisfying the
constraint $\int f(x)h(x)dx = const$ is of the form
\begin{equation}
f(x) = \exp\left[ \lambda_0 + \lambda h(x)\right].
\label{eq:S-general}
\end{equation}
The constants $\lambda_0$ and $\lambda$ are fixed by the
requirement of normalization for $f(x)$ and by the constraint.
Now, the condition
\begin{equation}
\langle z \rangle = z_0 = \frac{q - 1}{2 - q}\quad {\rm
where}\quad z = \ln \left[1 - (1 - q)\frac{x}{x_0} \right]
\label{eq:condT}
\end{equation}
results in
\begin{equation}
 f(z) = \frac{1}{z_0} \exp \left( - \frac{z}{z_0} \right)\, =\,
 \frac{1}{\left(1 + z_0\right)x_0}\left[ 1 +
 \frac{z_0}{1 + z_0}\frac{x}{x_0}\right]^{\frac{1 + z_0}{z_0
 }}\, =\, \frac{2 - q}{x_0} \left[ 1 - (1 - q)\frac{x}{x_0}
 \right]^{\frac{1}{1 - q}} , \label{eq:TfS}
\end{equation}
i.e., in a Tsallis distribution. For more detailed presentation
cf. \cite{S-T}.

\section{Conclusions}

We close with the following remark. It turns out that the
condition imposed by Eq. (\ref{eq:condT}) is natural for processes
described by Eq. (\ref{eq:L}), i.e., for multiplicative noise.
Notice the connection between the kind of noise in this process
and the condition imposed in the MaxEnt approach. For processes
described by additive noise, $dx/dt = \xi(t)$, one has exponential
distributions. The natural condition for them is that imposed on
the arithmetic mean, $\langle x\rangle = c + E(\xi)\tau $. For the
multiplicative noise, $dx/dt = x \gamma(t) $, one has a power law
distribution for which the natural condition is $\langle \ln
x\rangle = c + E(\gamma) \tau $, i.e., the geometrical mean (for
more details, cf. \cite{means}).

From the examples presented here it should be realized that the
widely discussed origin of Tsallis distribution as emerging from
Tsallis entropy, is by no means the only possibility. It also
arises from many nonthermal sources  without really resorting to
Tsallis entropy.


\begin{thebibliography}{10}

\bibitem{WWexample1} G. Wilk and Z. W\l odarczyk, \textit{Eur. Phys. J.}
                     \textbf{A40}, 299 (2009).

\bibitem{WWexample2} G. Wilk and Z. W\l odarczyk, \textit{Eur. Phys. J.}
                     \textbf{A48}, 162 (2012).

\bibitem{WWexample3} G. Wilk and Z. W\l odarczyk, \textit{Centr. Eur. J. Phys.} \textbf{10}, 568
                     (2012).

\bibitem{CYWWexample} C. Y. Wong and G. Wilk, \textit{Acta Phys. Pol.}
                      \textbf{B43}, 2047 (2012); \textit{Phys. Rev.}
                      \textbf{D87}, 114007 (2013).

\bibitem{Ss} C. Beck and E.G.D. Cohen, \textit{Physica} \textbf{A322}, 267
             (2003);
             F. Sattin, \textit{Eur. Phys. J.} \textbf{B49}, 219 (2006).

\bibitem{WW} G. Wilk, Z. W\l odarczyk, \textit{Phys. Rev. Lett.} \textbf{84}, 2770
             (2000) and \textit{Chaos Solitons Fractals} \textbf{13}, 581
             (2001).

\bibitem{BP} T.S. Bir\'o, A. Jakov\'ac, \textit{Phys. Rev. Lett.} {\textbf94},
             132302 (2005).

\bibitem{qnets} G. Wilk and Z. W\l odarczyk, \textit{Acta Phys. Pol.} \textbf{B35},
                871 (2004) and 2141 (2004); C. Tsallis, \textit{Eur. Phys. J.} \textbf{ST161},
                175 (2008); D.J.B. Soares, C. Tsallis, A.M. Mariz and L.R. da Silva,
                \textit{Europhys. Lett.} \textbf{70}, 70 (2005).

\bibitem{AT} C. Anteneodo and C. Tsallis, \textit{J. Math. Phys.} \textbf{44}, 5194 (2003).

\bibitem{NA49} C. Alt et al., \textit{Phys. Rev.} \textbf{C77}, 034906 (2008) and
               \textit{Phys. Rev.} \textbf{C77}, 024903 (2008);
               S. V. Afanasiev et al., \textit{Phys. Rev.} \textbf{C66}, 054902
               (2002).

\bibitem{CMS} V. Khachatryan et al. [CMS Collaboration], \textit{J. High Energy Phys.}
              \textbf{02}, 041 (2010) and \textit{Phys. Rev. Lett.} \textbf{105}, 022002
              (2010); S. Chatrchyan et al. [CMS Collaboration], \textit{J. High Energy Phys.}
              \textbf{08}, 086 (2011).

\bibitem{Wibig} T. Wibig, \textit{J. Phys.} \textbf{G37}, 115009 (2010).

\bibitem{RHIC} B. De, S. Bhattacharyya, G. Sau and S.K. Biswas, \textit{Int. J.
               Mod. Phys.} \textbf{E16}, 1687 (2007).

\bibitem{ATLAS} G. Aad et al. [ATLAS Collaboration], \textit{New J. Phys.} \textbf{13},
                053033 (2011).

\bibitem{PRC} G. Wilk and Z. W\l odarczyk, \textit{Phys. Rev.} \textbf{C79}, 054903
             (2009).

\bibitem{S-T} E. Rufeil Fiori and A. Plastino, \textit{Physica} \textbf{A392},
              1742 (2013).

\bibitem{means} A.Rostovtsev, {\it On a geometric mean and power-law statistical
                distributions}, arXiv:cond-mat/0507414
                [cond-mat.stat-mech].

\end{thebibliography}
\end{document}